\documentclass[12pt]{iopart}
\usepackage{amsmathMM}
\usepackage{iopams}  

\usepackage[dvips]{graphicx}
\usepackage[ansinew]{inputenc}
\usepackage{feynmf}
\usepackage{pst-grad,color}

\def\abs#1{\mathinner{\lvert#1\rvert}}
\newcommand{\up}{\uparrow}
\newcommand{\down}{\downarrow}

\begin{document}

\title[Crossover from adiabatic to sudden interaction quenches in the Hubbard model]{Crossover from adiabatic to sudden interaction quenches in the Hubbard model: Prethermalization and nonequilibrium dynamics}  

\author{Michael Moeckel$^{1,2}$, Stefan Kehrein$^1$}

\address{$^1$ Ludwig-Maximilians-Universit\"at M\"unchen, Department f\"ur Physik und Arnold-Sommerfeld-Center for Theoretical Physics, Theresienstra\ss{}e 37, D-80333 M\"unchen, Germany}
\address{$^2$ Max-Planck-Institut f\"ur Quantenoptik,  Hans-Kopfermann-Stra\ss{}e 1, D-85748 Garching, Germany}
\ead{\mailto{Michael.Moeckel@physik.lmu.de},\mailto{Stefan.Kehrein@physik.lmu.de}}

\begin{abstract}
The recent experimental implementation of condensed matter models in optical lattices has motivated research on their nonequilibrium behavior. Predictions on the dynamics of superconductors following a sudden quench of the pairing interaction have been made based on the effective BCS Hamiltonian; however, their experimental verification requires the preparation of a suitable excited state of the Hubbard model along a twofold constraint: (i) a sufficiently nonadiabatic ramping scheme is essential to excite the nonequilibrium dynamics, and (ii) overheating beyond the critical temperature of superconductivity must be avoided. For commonly discussed interaction ramps there is no clear separation of the corresponding energy scales. 
Here we show that the matching of both conditions is simplified by the intrinsic relaxation behavior of ultracold fermionic systems: For the particular example of a linear ramp we examine the transient regime of \emph{prethermalization} [M. Moeckel and S. Kehrein, Phys. Rev. Lett. {\bf 100}, 175702 (2008)] under the crossover from sudden to adiabatic switching using Keldysh perturbation theory. A real-time analysis of the momentum distribution exhibits a temporal separation of an early energy relaxation and its later thermalization by scattering events. For long but finite ramping times this separation can be large. In the prethermalization regime the momentum distribution resembles a zero temperature Fermi liquid as the energy inserted by the ramp remains located in high energy modes. Thus ultracold fermions prove robust to heating which simplifies the observation of nonequilibrium BCS dynamics in optical lattices. 
\end{abstract} 

\vspace{2pc}

\submitto{\NJP}

\section{Introduction}
Stimulated by the development of new experimental techniques in the field of ultracold atom gases and pump-probe laser spectroscopy, the simulation and implementation of established condensed matter model Hamiltonians in novel experimental setups and the study of their nonequilibrium behavior has florished recently \cite{Cazalilla2006, Manmana2007, Gangardt2007, Rigol2006A, Rigol2007,Kollath2007,Gritsev2007,Cazalilla2009, Eckstein2008,Rosch2008,Eckstein2009, Hackl2009}. 
In these systems precise and rapid control over parameters can be combined with a high temporal resolution for the observation of the many-body quantum dynamics. Thus, the switch of an interaction and the analysis of the subsequent response allows to address the interplay of particle correlations imposed by interactions and nonequilibrium initial conditions.  

For many systems a slow adiabatic change of a two-particle interaction may modify the ground state(s) of a system but no excitations to higher energy states occur. Therefore, adiabatic switching procedures allow to develop equilibrium properties of interacting systems from those of noninteracting ones. Due to the broad applicability of this approach they have been counted among the "basic notions" of condensed matter theory \cite{AndersonBNCMP}. 
The most prominent example is Landau's theory of a Fermi liquid \cite{Landau1957a}: Based on its main prerequisite, the adiabatic increase of the fermion interaction starting from the noninteracting Fermi gas, Landau could establish a one-to-one relation between the eigenstates of the noninteracting Fermi gas and approximate eigenstates of the interacting Fermi liquid. This link allows to carry over an approximate particle-like description to the interacting system and leads to the celebrated and imaginative concept of quasiparticles: reminiscent of the constituent physical fermions, albeit modified by a renormalization of their energies and masses, they represent the elementary excitations of the interacting system. Note that the quality of this approximation, i.e. the considerable time stability of low energy quasiparticles, is owed to phase space restrictions for elastic scattering of fermions around the Fermi energy: these restrictions, which emerge from the Pauli exclusion principle for fermions, prevent a fast decay of approximate quasiparticles into the true eigenstates of the interacting system and are the root of the universal applicability of Fermi liquid theory to many interacting fermionic systems. 

However, nonadiabatic switching, e.g. of the strength of trapping potentials or magnetic fields, of optical lattices or by applying ultrashort laser (pump) pulses may drive a system away even from its interacting ground state. Hence, fast parameter switching has shown to be a helpful tool to impose nonequilibrium initial conditions onto many-particle quantum systems. The limiting case of an instantaneous switch, called a quantum quench, provides a convenient starting point for theoretical investigations. Afterwards, the evolution of the excited many-body state can be followed in real time and different stages of the relaxation dynamics have been observed.

In fermionic systems, nonadiabatic switching of the interaction challenges the main prerequisite of Fermi liquid theory. Nonetheless, for not too strong sudden interaction quenches within the Fermi liquid phase of the Hubbard model, it has been predicted analytically \cite{Moeckel2008, Moeckel2009}, confirmed numerically \cite{Eckstein2009} and furtheron discussed \cite{Uhrig2009, Babadi2009} that during the initial stage of the relaxation a nonequilibrium description builds up which is strongly reminiscent of Fermi liquid theory. The same fermionic phase space restrictions which ensure the persistence of quasiparticles in equilibrium restrict a rapid relaxation of the momentum distribution in nonequilibrium. Hence, a transient regime of \emph{prethermalization} has been observed. Whereas, for instance, the total kinetic energy has already relaxed to its final value, the momentum distribution combines different features: a sharp step at the Fermi energy indicates zero temperature while deviations from its equilibrium shape represent the nonequilibrium nature of the transient state. 
Hence, during the first phase of the relaxation dynamics no substantial conversion of the initial excitation energy into heat can be observed and thermalization is deferred to a later time scale \cite{Moeckel2008}. 

In this paper we generalize these observations to finite switching times, assuming a linear ramping procedure. This introduces a new time scale, the \emph{ramp-up time} which is defined as the time duration on which the parameter change occurs. This ramp up scale can be used deliberately by finite-time 'ramp-up' protocols to prepare a system in particular (excited) states which are different from those reachable by sudden switching \cite{Duerr2004}. For instance, ramping on finite time scales has been suggested as a way to access the nonequilibrium regime of systems which are sensitive to heating effects since slower parameter changes reduce the excitation energy inserted into a system. 

In particular, this argument has been applied for the analysis of superconducting BCS systems in nonequilibrium. Theoretical studies have examined an interaction quench in the BCS Hamiltonian and predicted interesting nonequilibrium dynamics, e.g. oscillations of the order parameter in time \cite{Barankov2004,Yuzbashyan2005,Yuzbashyan2006b,Yuzbashyan2006, Warner2005, Barankov2006, Barankov2006A}.
However, the BCS interaction emerges from a local two-particle interaction between fermions which is the tunable switching parameter in experiments with ultracold fermions loaded onto an optical lattice. A realistic modelling of nonequilibrium dynamics in ultracold atomic gases therefore has to start with a quench of a spatially local interaction. Hence, BCS quench dynamics must be discussed against the background of a quenched Fermi liquid. Obviously, BCS dynamics can only be expected for temperatures below the critical temperature $T_c$ of superconductivity. 
A first approach based on a golden rule argument suggested that overheating can be avoided as energy scales of heating and of the nonadiabaticity requirement imposed on the ramp should separate \cite{Barankov2006A}. However, we will show in the sequel that this argument neglects contributions of high energy momentum modes which are relevant in commonly discussed ramping models. Therefore, the argument turns out inconclusive in the case of linear ramping. Fortunately, the problem of overheating is mitigated by two aspects: In another article published in this focus issue M. Eckstein and M. Kollar discuss the dependence of the excitation energy inserted by interaction ramping on the functional shape of the ramp \cite{EcksteinNJP2009}; this opens space for optimizing the ramping procedure.  

We address, on the other hand, the intrinsic robustness of fermionic systems at zero temperature to heating. Even for ramping scenarios which do not lead to a clear energy separation of the relevant BCS energy scale and the effective temperature after heating we find a \emph{temporal} separation of the characteristic time scales of BCS dynamics and thermalization, respectively. Our observation mirrors the notorious problem of cooling a gas of cold fermionic atoms to even lower temperatures and views a current technical limitation as a desirable feature for future experiments: On the one hand, the Pauli principle suppresses s-wave scattering between the fermions. This hinders the equilibration of the gas at a lower temperature after high energy particles have been lost by evaporation and poses a great challenge to experimentalists on their way to reach a 'zero temperature' Fermi gas. On the other hand, we will show that the same mechanism prevents a rapid thermalization of excitations of high energy modes in a fermionic system which has been previously cooled down to zero temperature. Then a long-lasting transient state of prethermalization emerges during which no heating occurs. This behavior of cold fermionic many-particle systems will simplify the observation of nonequilibrium BCS dynamics during a long transient time window before thermalization sets in; since it roots in Fermi degeneracy caused by the Pauli principle it is independent of any details of the applied ramp. 

In this work we perform a careful investigation of the crossover from adiabatic to instantaneous linear switching procedures on the grounds of second order Keldysh perturbation theory. We follow the evolution of the momentum distribution function for \emph{original} fermions \footnote{Note that we analyze the nonequilibrium properties of the interacting system in terms of the unrenormalized eigenmodes of the noninteracting system. This differs from other approaches which assume the existence of a quasiparticle picture and discuss directly the behavior of nonequilibrium \emph{quasiparticle} distributions. } after a linear interaction ramp of finite duration $t_R$ and discuss the kinetic energy in a long-time limit of the second order result. We observe that the corresponding effective temperature is of the same order of magnitude as the critical temperature $T_c$. Hence overheating cannot be excluded by an argument based on energy scale separation. However, we still observe prethermalization behavior of the Fermi gas; therefore this temperature is only reached after a long transient regime. Since in this regime the momentum distribution resembles closely a zero-temperature Fermi liquid we conclude that the observation of nonequilibrium BCS dynamics can be expected \emph{before} overheating wipes out all BCS signatures.

\section{Hubbard model with time dependent interaction}
For an analysis of many-body quantum dynamics we implement Fermi liquid theory microscopically by the Hubbard Hamiltonian $H^{\rm HM} =H^{\rm HM}_{\rm kin}+H^{\rm HM}_{\rm int}$. It provides a paradigmatic description for correlated fermions on a lattice; it includes spin and implements the Pauli principle by restricting to a local state space of dimension four. 
The itinerant properties of the fermions are contained in the kinetic part of the Hamiltonian and described as coherent hopping processes between neighboring lattice sites with a hopping matrix element $h$. The chemical potential $\mu$ is fixed to ensure half filling. 
\begin{equation}
H^{\rm HM}_{\rm kin} = 
-h\sum_{\langle i,j\rangle \sigma} \ c^{\dagger}_{i\sigma}c_{j\sigma} 
-\mu \sum_j \left( n_{j\up}+n_{j\down} \right)
\stackrel{\rm FT}=  
\sum_{k \in \mathcal{K},\sigma \in \lbrace \up, \down \rbrace } \left(\epsilon_k -\mu \right) \ {c^{\dagger}_{k\sigma}c_{k\sigma}} 
\end{equation}
A momentum space representation is obtained by a Fourier transform. The dispersion relation $\epsilon_{k}$ depends on the details of the lattice. 
For convenience, explicit numerical calculations of the momentum distribution are performed on a hypercubic lattice in the limit of infinite dimensions ($d \rightarrow \infty$) for which a Gaussian density of states is applicable \cite{Vollhardt1992}. 
\begin{equation}
\label{GaussianDoS}
\rho(\epsilon)=\exp\left(-(\epsilon/h^{*})^{2}/2\right)/\sqrt{2\pi}h^{*}
\end{equation}
$h^*$ is linked to the hopping matrix element by dimensional scaling $h \longrightarrow \frac{h^*}{\sqrt{2d}}$ \cite{Vollhardt1992}. For convenience, we set $\hbar=1$ and $k_B=1$.

Correlations between the lattice fermions are modeled by a local two-particle interaction $U$ acting between fermions of different spin with $n_{j\sigma}$ denoting the local occupation 
\begin{equation}
\label{INTRO_HM_HaInt}
{H}^{\rm HM}_{\rm int} = U(t) \sum_j \left(n_{j \up}-\frac 12\right)\left(n_{j\down}-\frac 12\right) 
\end{equation}
Since our interest is in Fermi liquid properties of the model, we restrict to a regime of weak interaction strength ($U(t) \ll h^* \ \forall t$). This rules out the observation of any features of the Mott-Hubbard phase transition. In consequence, the hopping matrix element is a natural choice for the energy unit ($h^*\equiv 1$).

To study switch-on processes we assume that the bare interaction $U(t)$ is a time dependent parameter. For simplicity, linear ramps of the interaction are considered which are characterized by a ramp-up time scale $t_R$.
\begin{equation}
U(t) = U \left\lbrace \begin{array}{cl}
0 & t\leq 0 \\
t/t_R   & 0<t<t_R  \\
1 & t>t_R 
\end{array} \right.
\label{DefLinearRamping}
\end{equation}
In the sequel we discuss the effects of variations of the parameter $t_R$ onto the real-time evolution of the momentum distribution. We assess the dependence of the heating on the ramp-up time $t_R$ by analyzing the $t_R$-dependence of the kinetic energy for deviations from adiabatic switching procedures.

\section{Keldysh perturbation theory for time dependent Hubbard model}

Perturbation theory for the Hubbard model has been set up previously by various authors \cite{Noack2003}. Here a real-time Keldysh approach following Rammer \cite{Rammer_NEQQFT} allows to extract the time dependent behavior of the momentum distribution function $N^{\rm NEQ}_k(t)$ from the Keldysh component Greens function ${ G^K}(x, t, \sigma_z, x, t, \sigma_z)$. Both functions are evaluated in second order of the time dependent local two-particle interaction and in the initial representation of noninteracting fermions. The later implies that interaction effects become visible only by a redistribution of occupation among the unrenormalized momentum modes. 
\begin{equation}
\label{KeldyshMDFwithKGFdef}
N^{\rm NEQ}_k(t) = -i \sum_{\sigma_z} \int dx e^{ikx} { G^K}(x, t, \sigma_z, x, t, \sigma_z)
\end{equation}
Since first-order diagrams vanish due to symmetry, only a single second order diagram has to be evaluated for a consistent second order result of the momentum distribution function. Details of the calculation can be found in \ref{AppendixKeldysh} or in \cite{MoeckelPhD}. 

The second order result describes the dynamics of the momentum distribution during a first stage of its relaxation and exhibits the early relaxation of the kinetic energy. Higher order corrections, however, induce different behavior on a second time scale. We describe this second stage by means of the quantum Boltzmann equation, taking the formal long-time limit of the second order Keldysh perturbative calculation as an initial condition. It will lead to a thermalization of the momentum distribution. In section \ref{SectionEffectiveHeating} we will justify this treatment, explain the separation of both relaxation scales as a consequence of fermionic phase space restrictions and discuss the characteristics of the transient regime of prethermalization.

\section{Evaluation of the momentum distribution}
We have evaluated the time dependent momentum distribution function of original fermions in the regime $t>t_R $ for arbitrary ramp-up times $t_R$. Within the validity of second order time-dependent perturbation theory, i.e. for not too large times $t$, the momentum distribution function reads
\begin{multline}
\Delta N^{\rm NEQ}_k(t,t_R) =
-U^2 \int dE \ J_k(E;n) \ \bigg\lbrace \frac 1{t_R^2} \frac{2\big(1-\cos((E-\epsilon_k) t_R)\big)}{(E-\epsilon_k)^4} + \frac 1{(E-\epsilon_k)^2}- 
\label{KeldyshMDFresultKSpace}\\
- \frac 2{t_R} \frac{ \sin((E-\epsilon_k)t_R)\cos((E-\epsilon_k)t) + \sin((E-\epsilon_k)t)\big(1-\cos((E-\epsilon_k)t_R)\big)}{(E-\epsilon_k)^3} \bigg\rbrace 
\end{multline}
$J_k(E;n)$ represents a fermionic phase space factor which is discussed in more detail in \ref{AppendixJ}. Here we note that in the case of $\epsilon_k > 0$ it is nonvanishing only for negative energies $E$, in the case of $\epsilon_K < 0 $ for positive ones. Moreover, it asymptotically approaches a quadratic energy dependence for small absolute values of $E$, i.e. $\lim_{E \rightarrow 0}  J(E;n) \sim E^2$. Together, these observations imply that all formal energy divergences in equation (\ref{KeldyshMDFresultKSpace}) are cut off or regularized by the phase space factor.

\subsection{No secular terms}
\label{Keldysh_SecularTerms}
A second glance at the result (\ref{KeldyshMDFresultKSpace}) shows that --contrary to the Keldysh Greens function itself (see appendix \ref{AppendixKeldysh})-- the momentum distribution function does not exhibit secular terms which are proportional to time $t$ and typically would restrict the validity of a perturbative approach to short time regimes. This advantage is a consequence of the simplicity of the number operator and of the linear ramping assumption (\ref{DefLinearRamping}).  

\subsection{Limiting cases for the linear ramping}
For the correction to the momentum distribution two limiting cases have been previously discussed based on a flow equation calculation \cite{Moeckel2008, Moeckel2009}. Both limits are reproduced by Keldysh perturbation theory. 

\subsubsection{Adiabatic limit}
The adiabatic limit is given by an arbitrary slow linear increase of the interaction strength and corresponds to $t_R \rightarrow \infty$. Then only the $t_R$-independent term $(E-\epsilon_k)^{-2}$ contributes.
\begin{eqnarray}
\lim_{t_R  \rightarrow \infty} \Delta N^{\rm NEQ}_k(t,t_R ) &=&
-U^2 \int dE \ J_k(E;n) \frac 1{(E-\epsilon_k)^2}
=: \Delta N^{\rm EQU}_k
\label{AdiabaticLimitN}
\end{eqnarray}
This is a truly stationary state. It equals the second order equilibrium result $\Delta N^{\rm EQU}_k$ for a microscopic implementation of Landau's Fermi liquid theory.  

\subsubsection{Quench limit}
The limit of a {q}uantum {q}uench, i.e. a sudden switch-on of the interaction, is obtained for $t_R \rightarrow 0$ and represents the maximally nonequilibrium case for any linear ramping procedure. Replacing \mbox{$1-\cos((E-\epsilon_k) t_R)= 2 \sin^2((E-\epsilon_k) T/2)$} the correction to the momentum distribution shows a nontrivial time evolution which incorporates the initial build-up of correlations.
\begin{equation}
\label{KeldyshMDFres}
\Delta N^{\rm qq}_k(t):= \lim_{t_R \rightarrow 0} \Delta N^{\rm NEQ}_k(t,t_R) =
-U^2 \int dE \ J_k(E;n) \frac{2\big(1-\cos((E-\epsilon_k) t)\big)}{(E-\epsilon_k)^2} 
\end{equation}
The nonequilibrium nature of this distribution is best analyzed in the long-time limit of this second order result. It is defined for any time dependent function $f(t)$ by 
$$
\overline{f(t)} = \lim_{t \rightarrow \infty} \frac 1t \int_0^t dt' f(t')
$$ 
Note that the notation $\overline{f(t)}$ intends to resemble the \emph{generic} time dependence of the function $f(t)$ but not a possible time dependence of the long-time average. 
In this limit a characteristic mismatch $\mu(\epsilon_k)$ between nonequilibrium distribution and a corresponding equilibrium one --computed for the same interaction strength $U$-- has been found and will be discussed in great detail later.

\section{Discussion of the crossover from instantaneous to adiabatic switching}
We discuss the crossover from instantaneous to adiabatic switching by analyzing two observables which show significantly different behavior, namely
\begin{itemize}
\item[1.] the time dependent momentum distribution function $N^{\rm NEQ}_k(t,t_R )$ and
\item[2.] the kinetic energy $E_{\rm kin}$ in a formal long-time limit of a perturbative result
\end{itemize}
The observed differences illustrate and resolve prior confusion on the relevant energy scales of fermionic systems in nonequilibrium. 

\subsection{Time dependent momentum distribution function}
After the interaction has been switched on, correlations between particles develop. In the following we observe their impact on the momentum distribution, leading to a redistribution of occupation among the momentum modes with time. Features of the long-time limit of this dynamics are discussed in a second step.

\subsubsection{Correlation buildup for arbitrary ramping times.}
\begin{figure}
 \begin{center}
      \includegraphics[width=177mm]{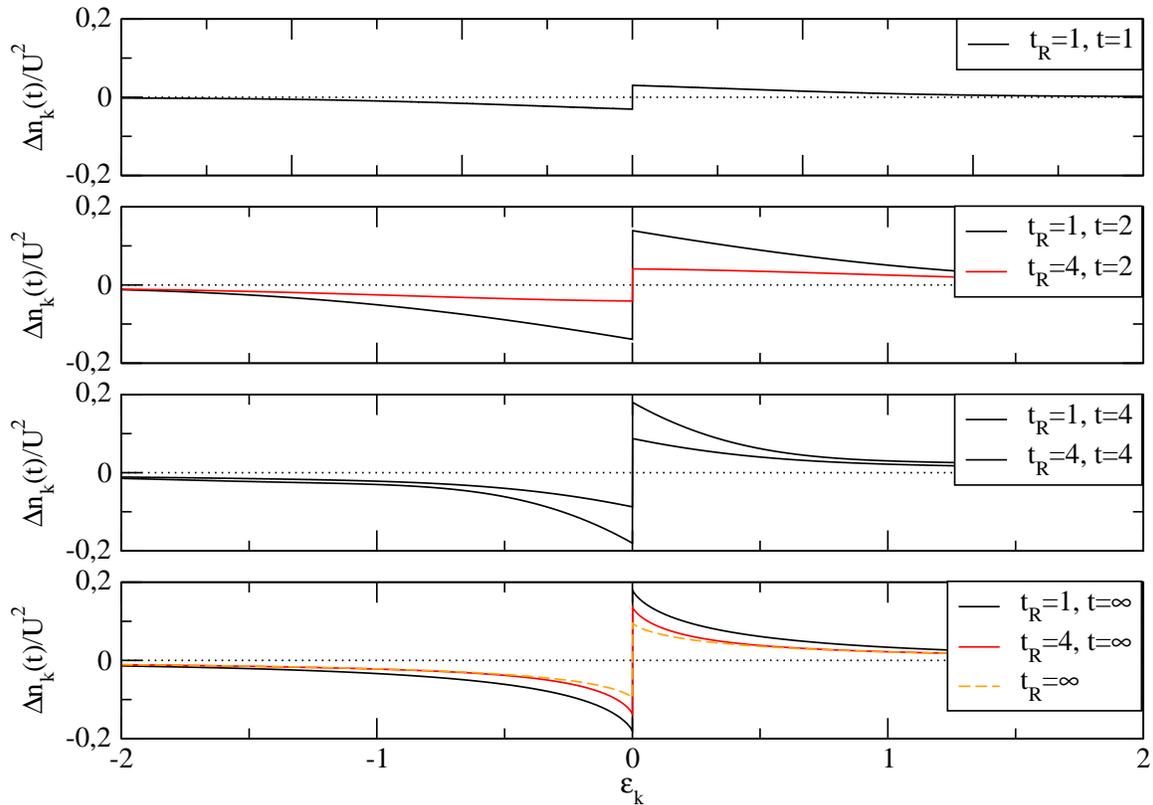}  
    \caption[]{Build-up of an interacting Fermi liquid description after fast ($t_R=1$), intermediate ($t_R=4$) and adiabatic ($t_R \rightarrow \infty$) ramping.}
      \label{Fig1}
 \end{center}
\end{figure}

The initial buildup of correlations is mirrored by a reshape of the momentum distribution. In figure \ref{Fig1} we picture this process for five time steps, comparing three ramping scenarios: (1) a fast ramp-up of the interaction (black line for $t_R=0$), (2) an intermediate ramping speed (red line, $t_R=4$) and (3) the adiabatic limit ($t_R\rightarrow \infty$). 
In the first case the evolution of the momentum distribution can be clearly followed: For small times, 1/t-oscillations depict a rapid redistribution of the initial occupation (at $t=1$) into many high-energy modes. With increasing time, occupation is transferred back towards the Fermi energy such that in the long-time limit changes to the momentum distribution are most relevant around the Fermi energy.

For intermediate switching times, the process of redistribution of momentum mode occupation occurs in a decelerated way. Since we have assumed $t>t_R$ in the derivation of (\ref{KeldyshMDFresultKSpace}) the second time slice $(t_R=4, t=2)$ does not include possible transient corrections which might change the short-time dynamics. 

A comparison of all three ramp-up scenarios in the long-time limit shows that the absolute change in the momentum distribution is reduced with increasing ramp-up time $t_R$. Most clearly, this is seen at the Fermi energy. 

\subsubsection{Mismatch function.} 
A quantitative measure for the deviations of the long time limit of the nonequilibrium momentum distribution ($t_R < \infty$) from the equilibrium one ($t_R\rightarrow \infty$) is given by the mismatch function
$$
\mu(\epsilon_k,t_R) =  \frac{\overline{\Delta N^{\rm NEQ}_k(t,t_R)}}{\Delta N^{\rm EQU}_k(t_R)}
$$
While for adiabatic switching on of the interaction $\mu(\epsilon_k, t_R\rightarrow \infty)\equiv 1$ is trivial for all values of $\epsilon_k$, for finite ramp-up times deviations become visible around the Fermi energy. Within a finite environment of approximate radius $\epsilon_R \sim 2/t_R$ the values of $\mu(\epsilon_k, t_R)$ increases up to a maximum value of two which is reached in the limiting case of a sudden interaction quench. This confirms previous observations based on a flow equations approach \cite{Moeckel2008, Moeckel2009}. 

\begin{figure}
 \begin{center}
      \includegraphics[width=150mm]{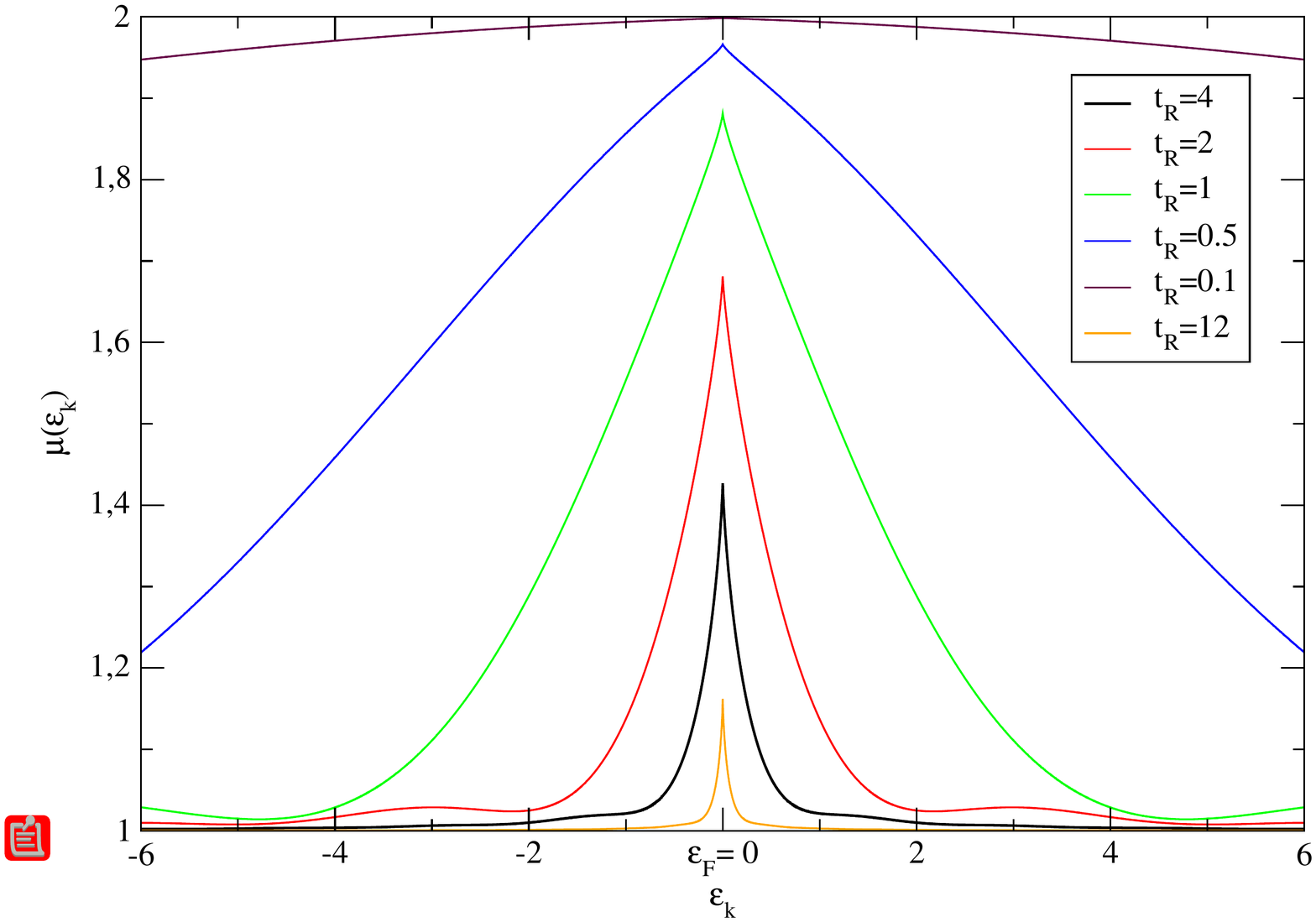}  
    \caption[]{The mismatch function $\mu(\epsilon)$ is plotted for different ramp-up times. Deviations from the value $\mu(\epsilon) = 1$ depict the nonequilibrium nature of the transient state which is described by the long-time limit of a second order perturbative calculation. Those are bounded by an upper value of $\mu(\epsilon_F) = 2$ in the limit of a quantum quench. For small corrections to the adiabatic limit, i.e. for long ramp-up times, they are sharply peaked around the Fermi energy.}
      \label{Fig2}
 \end{center}
\end{figure}

Fig. \ref{Fig2} shows numerical computations of  $\mu(\epsilon_k)$ for different ramp-up times. While for strictly sudden switching the maximum value of two is assumed for all energies, even for very fast switching ($t_R=0.1$) a reduction of this value at the band edges (bandwidth D=6) is noticeable. On the other hand, even slight deviations from adiabatic switching leave an observable 'spike' around the Fermi energy in the mismatch function $\mu(\epsilon_k, t_R \gg 1)$. This proofs that around the Fermi edge the momentum distribution is highly sensitive even for small departures from the adiabatic limit.   

\begin{figure}
 \begin{center}
      \includegraphics[width=100mm]{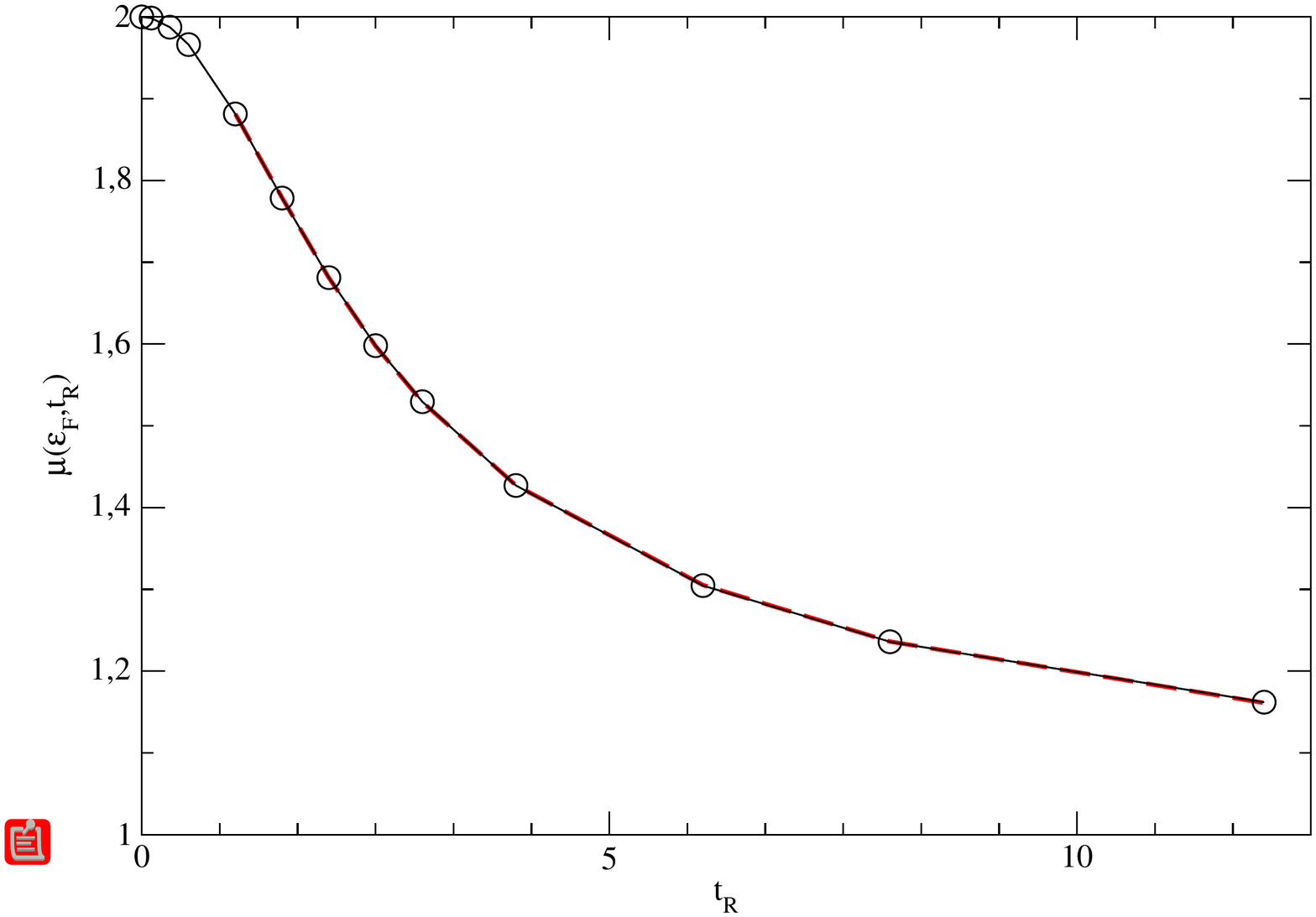}  
    \caption[]{Numerically calculated values of the mismatch function at the Fermi energy  (circles) are given for different ramp-up times $t_R$. Within a wide range of ramping times, correction to the adiabatic limit ($t_R\rightarrow \infty$) can be fitted by $\mu(\epsilon_F,t_R ) \approx 1.03 + 1.76 1/t_R  - 0.92 1/t_R ^2 + 0.62 1/t_R ^3$ ($\chi^2\approx 10^{-6}$, 9 data points were fitted); this agrees with the asymptotic expectation $\mu(\epsilon_F,t_R \rightarrow \infty)=1$. Close to a sudden switch there is saturation at the value $\mu(\epsilon_F,t_R \rightarrow 0)=2$. The crossover between both regimes can be estimated as $\tilde{t_R } \approx 1/h^*\equiv 1$ where the inverse power-law fitting has broken down.}
      \label{Fig3}
 \end{center}
\end{figure}

Deviations from the adiabatic behavior are most pronounced at the Fermi energy. Hence the (numerical resolvable) peak height of the mismatch function at the Fermi energy $\mu(\epsilon_F,t_R)$ defines a simple measure for the adiabatic character of the momentum distribution. Fig \ref{Fig3} displays its dependence on the ramp-up time for some values of $t_R$. 

\subsubsection{Intrinsic energy scales of the momentum distribution.} 
\label{IESMDF}
From the peak width in figure \ref{Fig2} one may be tempted to read off the appearance of a new energy scale $\epsilon_R \sim 1/t_R$ in the nonequilibrium momentum distribution. It describes the extension of an environment around the Fermi energy $\epsilon_F \equiv 0$ where significant deviations from equilibrium can be observed. Note that it is not equivalent to a scale set by an effective temperature since the momentum distribution still exhibits a discontinuity at the Fermi energy which indicates zero-temperature behavior.

The same scale can be extracted from a formal analysis of the momentum distribution. 
Since the sinusoidal functions in (\ref{KeldyshMDFresultKSpace}) can be discussed in two limiting cases known as the 'rapid oscillation' and the 'small angle' regime, two energy scales  $\epsilon_R = 1/t_R$ and $1/t$ emerge. However, only the first one is relevant in the long-time limit. This motivates the discussion of two extremal cases of the internal energy integration in (\ref{KeldyshMDFresultKSpace}): $\abs{E-\epsilon_k} \gg 1/t_R$ (a) and $\abs{E-\epsilon_k} \ll 1/t_R$ (b). 
Due to the properties of the phase space factor $J_k(E,n)$ (cf. \ref{AppendixJ}) these restrictions give rise to an analogous separation of regimes for the external energies $\epsilon_k \gtrless 1/t_R$. 

In the first case, the rapid oscillations and the small $2/t_R^2(E-\epsilon_k)^4$ contributions to the integral can be neglected.
\begin{equation}
\left.\Delta N^{\rm NEQ}_k(t,t_R) \right|_{\abs{E-\epsilon_k} \gg 1/t_R} = 
- U^2 \left. \int dE \frac{J_k(E,n)}{(E-\epsilon_k)^2} \right|_{\abs{E-\epsilon_k} \gg 1/t_R} \approx 
\left.\Delta N^{\rm EQU}_k  \right|_{\abs{\epsilon_k} \gg 1/t_R}
\end{equation}
For finite ramp-up times and large values of $\abs{\epsilon_k}$ the condition (a) is satisfied for all values of $E$. Hence far away from the Fermi energy the nonequilibrium momentum distribution equals the adiabatic one (\ref{AdiabaticLimitN}). 

In the second case we consider the long-time limit such that the time dependent arguments average out. Since now $\abs{E-\epsilon_k} < 1/t_R$ the main contributions to the energy integral in (\ref{KeldyshMDFresultKSpace}) originate from the term
$$
\left. \int dE J_k(E,n) \left[\frac 2{t_R^2} \frac{\left(1-\cos\left((E-\epsilon_k)t_R\right) \right)}{(E-\epsilon_k)^4} \right|_{\abs{E-\epsilon_k} < 1/t_R} + \frac  {J_k(E,n) } {(E-\epsilon_k)^2} \right]
$$
In the case of rapid switching the condition $\abs{E-\epsilon_k} < 1/t_R$ can be satisfied both for a large range of energies $\epsilon_k$ and, at the Fermi energy, for a large energy interval in $E$. This implies substantial deviations from the equilibrium behavior in an environment around the Fermi energy. There it leads to the increase of the mismatch function up to a value of two.
$$
\left.\Delta N^{\rm NEQ}_k(t,t_R) \right|_{\abs{\epsilon_k} \ll 1/t_R} 
\approx {\bf 2}\left. \Delta N^{\rm EQU}_k \right|_{\abs{\epsilon_k} \ll 1/t_R}
$$
For slower ramping times the restriction cuts off the internal energy integration such that the maximum value of two is not reached any more.

\section{Effective heating for deviations from adiabatic switching}
\label{SectionEffectiveHeating}
We start the discussion of the kinetic energy by considering small nonadiabatic corrections to an adiabatic switching procedure, i.e. we focus on large ramp-up times. While the energy scale $\epsilon_R = 1/t_R$ allows to discuss energetically well-separated limiting cases of the momentum distribution, it leads to wrong expectations when it is carried over naively to the kinetic energy. We will discuss this observation and its consequences in the following sections.

For the purpose of ramp-up  dynamics, the kinetic energy $\Delta E^{\rm KIN}(t_R)$ is defined as the excess of kinetic energy with respect to the corresponding interacting ground state. 
\begin{equation}
\Delta E^{\rm KIN}(t_R) :=  \int_{-\infty}^{\infty} d\epsilon_k \ \epsilon_k \ \delta n_k :=
\int_{-\infty}^{\infty} d\epsilon_k \ \epsilon_k 
  \left[ \ \overline{\Delta N^{\rm NEQ}_k(t,t_R)}-\Delta N^{\rm EQU}_k \right]
  \label{DefKinEnergy}
\end{equation}
Note that this quantity provides an integrative view on the implications of non-adiabatic switching; it includes energy contributions both from the pronounced nonequilibrium regime $\abs{\epsilon_k} \ll 1/t_R$ and from a large crossover region $\abs{\epsilon_k} \gtrsim 1/t_R$ of higher energy modes.

\subsection{Full numerical solution of the kinetic energy}
\begin{figure}
 \begin{center}
      \includegraphics[width=150mm]{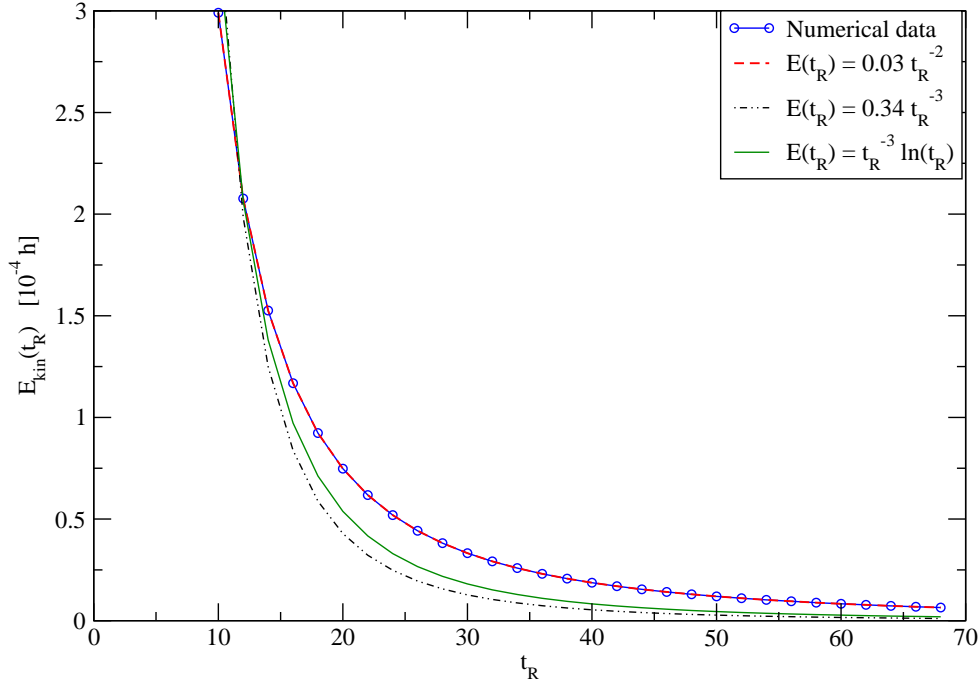}  
    \caption[]{The kinetic energy is evaluated numerically for large ramping times (circles). Fits for different power-law behavior are given. Obviously, only the inverse quadratic fit $\Delta E^{\rm KIN}(t_R)\sim 1/{t_R^2}$ agrees well with the data. }
      \label{Fig4}
 \end{center}
\end{figure}

A numerical integration of (\ref{DefKinEnergy}) after insertion of (\ref{KeldyshMDFresultKSpace}) exhibits the dependence of the kinetic energy $\Delta E^{\rm KIN}(t_R)$ on the ramp-up time (see figure \ref{Fig4}). 
For small corrections to the adiabatic limit, i.e. for long ramp-up times, this behavior agrees with expectations formulated on thermodynamic grounds \cite{PolkovnikovGritsev2008}.
\begin{equation}
\Delta E^{\rm KIN}(t_R)\sim \frac{U^2}{t_R^2}
\label{EtempDEP}
\end{equation}
The quadratic behavior can be attributed to the contributions of high energy modes with energies larger than a cut-off set by $1/t_R$. This can be seen if one deliberately cuts off these contributions in the numerical result: then a different dependence of the kinetic energy on the ramping $\Delta \tilde{E}^{\rm KIN}(t_R)\sim U^2 t_R^{-3}$ is rediscovered which has been published previously \cite{Barankov2006A, EcksteinNJP2009}. This indicates that the excitation energy per momentum mode $\epsilon_k  \delta n_k$ remains of order of the average $\Delta E^{\rm KIN}/D$ even for $\abs{\epsilon_k} \gg 1/t_R$ where $D$ represents the finite bandwidth. The additional power $1/t_R$ in  $\Delta \tilde{E}^{\rm KIN}(t_R)$ then merely reflects the arbitrarily chosen integration limits introduced by the cut-off.

\begin{figure}
 \begin{center}
      \includegraphics[width=120mm]{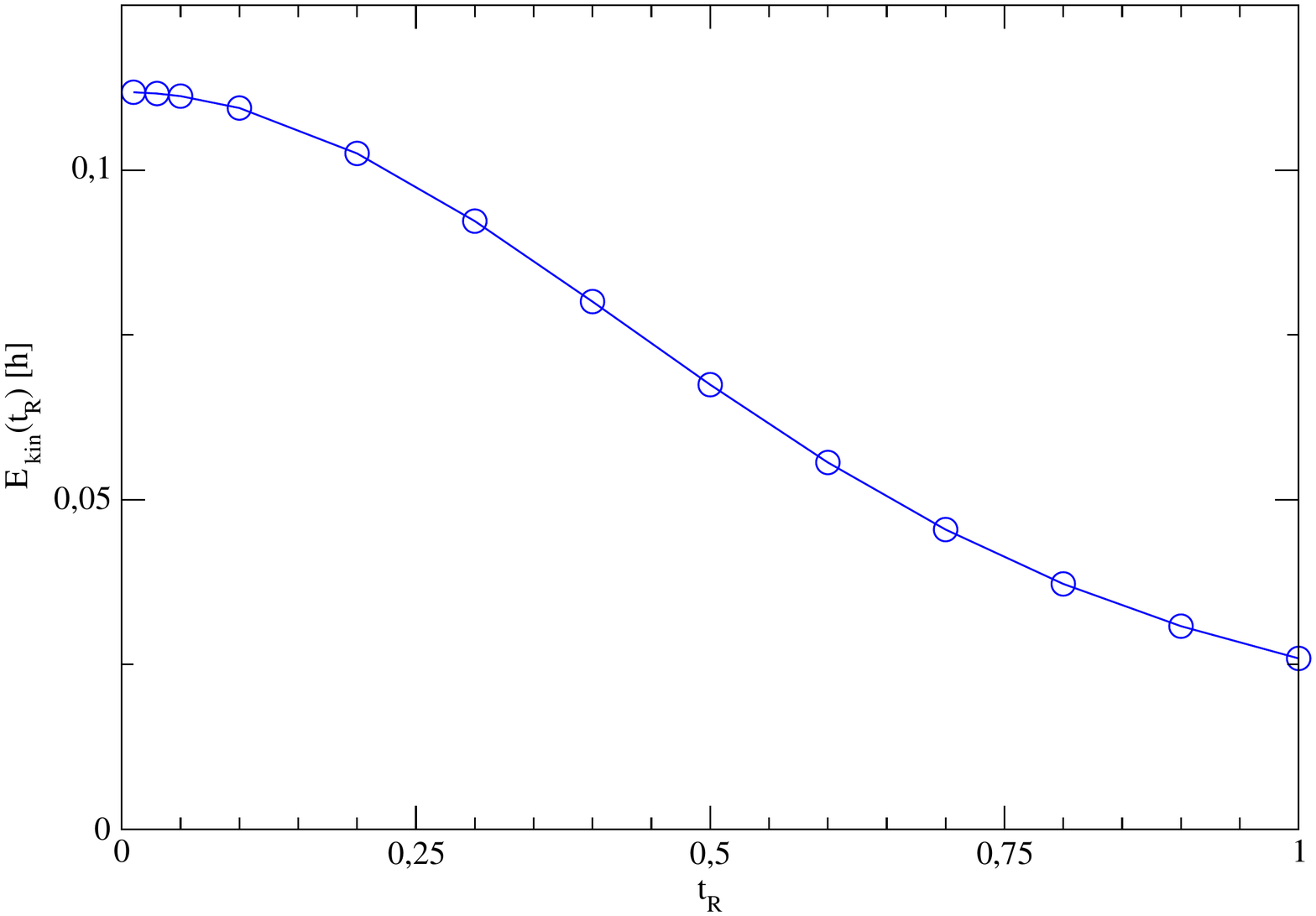}  
    \caption[]{The kinetic energy in the limit of fast ramp-up times saturates for a sudden quantum quench. }
      \label{Fig6}
 \end{center}
\end{figure}

In the opposite limit of a rapid switch the kinetic energy saturates at the value found for the quantum quench scenario (cf. figure \ref{Fig6}).

\subsection{Fermi's golden rule analysis for a ramping model}
The same observation can be made analytically in the ramping model defined by
$\lambda(t,t_R) = \lambda_0 \Theta(0)(1-e^{-t/t_R})$ which has been discussed before \cite{Barankov2006A}. In the following we will calculate the excitation energy according to Fermi's golden rule.
\begin{equation}
\Delta E^{\rm FGR} = \sum_{0<\omega<\Lambda} \omega \ \abs{\lambda(\omega)}^2 \ \underset{N(\omega)}{\underbrace{\left[ \sum_{1234}n_1 n_2 (1-n_3) (1-n_4) \delta_{1234} \right]}}
\label{BLkinE}
\end{equation}
We denote the phase space factor of two-particle scattering in the presence of a filled Fermi sea by $N(\omega)$ and analyze the role of an energy cut-off $\Lambda$.  

Fourier transforming the time dependent interaction reads  $\lambda(\omega,t_R) =i\lambda_0/\omega(1 -i\omega t_R)$ or  $\abs{\lambda(\omega,t_R)}^2 =\abs{\lambda_0}^2/\omega^2(1 +\omega^2 t_R^2)$. In a small frequency limit, 
the phase space factor for a Fermi liquid can be approximated by $N(\omega)\approx(\rho_F^4/p_F^4)\omega^3$ such that
\begin{multline}
\Delta E^{\rm FGR} = \int_{0}^{\Lambda} d\omega \ \frac{\abs{\lambda_0}^2}{\omega^2(1 +\omega^2 t_R^2)} \frac{\rho_F^4}{p_F^4}\omega^4 =
\frac {\rho_F^4}{p_F^4} \frac{\abs{\lambda_0}^2}{t_R^3} \int_{0}^{\Lambda t_R}
d(\omega t_R) \frac{(\omega t_R)^2}{1 +(\omega t_R)^2} = \\
= \frac {\rho_F^4}{p_F^4} \abs{\lambda_0}^2\frac{\Lambda t_R - \arctan(\Lambda t_R)}{t_R^3} 
\end{multline}
Now one easily observes that the dependence of the excitation energy on the ramping time is a direct consequence of the applied cut-off. 
Firstly, the pronounced low energy approach taken by Barankov and Levitov cuts off all contributions on the scale $\Lambda = 1/t_R$, i.e. assuming that for all relevant energy modes $\omega t_R  \ll 1$. This corresponds to an approximation of the Fourier transform by its low frequency asymptotic behavior $\lambda(\omega,t_R) \approx \lim_{\omega \rightarrow 0} \lambda(\omega,t_R) =i\lambda_0/\omega$. 
Since this approximation mirrors the Fourier transform of a sudden switch $\mathcal{F}[\Theta(t)](\omega)=i/\omega$ all details of the ramping model are lost. The resulting power-law behavior is $\Delta E^{\rm FGR}(t_R) \sim t_R^{-3}$. 

If one, on the other hand, one chooses a cut-off which is independent of the ramping time ($\Lambda \neq \Lambda(1/t_R)$) but small enough such that the low frequency approximation of the phase space still holds (this is possible for small corrections to adiabatic ramping) one observes the different behavior $\Delta E^{\rm FGR}(t_R) \sim t_R^{-2}$. This corresponds to replacing the Fourier transform of the switching model by its high frequency asymptotic limit $\lim_{\omega \rightarrow \infty} \lambda(\omega,t_R) =-\lambda_0/\omega^2 t_R$ and includes excitations which are characteristic for the particular ramping scheme. This approximation agrees well with our numerical calculations for linear ramping and is in full agreement with the general analysis provided in \cite{EcksteinNJP2009}. 
Since finite time ramping remains in all cases linked to a continuous increase of the interaction strength on the chosen time scale we expect that high frequency excitations will be present in many experimental implementations.  
However, in the following we will show that the influence of high frequency excitations on the low energy behavior can be delayed by a relaxation bottleneck. Independent of the applied ramping procedure the latter accounts for the emergence of a transient regime characterized by prethermalization. This defers the problem of overheating to late times of the relaxation dynamics.   

\subsection{Prethermalization for arbitrary switching times}
As we have already mentioned, the observation of a prethermalization regime has been first reported in the case of a sudden interaction quench in the Hubbard model. Here we have discussed the behavior of the momentum distribution function and of the kinetic energy on a first time scale set by second order perturbation theory $t_{\rm PT} \sim \rho_F U^{-2}$. For this time regime and in the case of finite switching times $ t_R \ll t_{\rm PT}$ we have observed that the additional kinetic energy remains smaller than the one inserted under a quantum quench; moreover, that the momentum distribution function retains a sharp step and differs only quantitatively but not qualitatively from that one of the quantum quench. In consequence, we note that the phase space restrictions for scattering processes are not overcome by finite ramping times. On the contrary, the nonequilibrium character of the momentum distribution is less pronounced which makes those restrictions more severe. This can be seen when the scattering integral $\mathcal{I}_k[N^{\rm QP}_k(t)]$ of a quantum Boltzmann equation
\begin{equation}
\frac {\partial N_k(t)}{\partial t} = \mathcal{I}_k[N_k(t)]
\label{QBE}
\end{equation}
is discussed. Note that the scattering integral vanishes for Fermi-Dirac distribution functions at arbitrary temperature and that the quantum Boltzmann equation provides an effective description for the evolution of a \emph{quasiparticle} distribution function. Quasiparticles are approximately noninteracting degrees of freedom which absorb most of the interaction effects present among the physical fermions. Therefore we construct a rough approximation to the quasiparticle mapping by the linear extrapolation 
$$
N_k^{\rm QP} =   N_k - {\Delta N_k^{\rm EQU}} 
$$
which strictly holds in the adiabatic limit. In the nonequilibrium case it provides an approximate representation of the quasiparticle momentum distribution. 
$$
\Delta N_k^{\rm NEQ:QP} =   \Delta N_k^{\rm NEQ} - {\Delta N_k^{\rm EQU}} 
$$
Note that $\Delta N_k^{\rm NEQ:QP} \sim \Or(U^2)$. Since around the Fermi energy $ \abs{\overline{\Delta N_k^{\rm NEQ}}} > \abs{{\Delta N_k^{\rm EQU}}} $ the second order perturbative long-time limit of the quasiparticle distribution shows a reduction of the quasiparticle residue and open phase space for quasiparticle scattering. Hence the scattering integral, initialized with this long-time limiting state,  does not vanish but gives rise to a subsequent Boltzmann dynamics of the momentum distribution (\ref{QBE}). Since its fixed point is a thermal state, thermalization can be predicted. The corresponding time scale can be extracted from a linearization of the scattering integral w.r.t. $\Delta N_k^{\rm NEQ:QP}$ around the zero-temperature initial state. At the Fermi energy the linearized scattering integral can be expressed in terms of the long-time limit of the kinetic energy $\overline{\Delta E^{\rm KIN}}$ (\ref{DefKinEnergy}) after the initial energy relaxation has been accomplished. \footnote{The linearization of the scattering integral around the zero temperature initial state is valid at the onset of thermalization. With increasing temperature, however, the linearization must be performed around a Fermi-Dirac distribution of nonzero temperature. This leads to a reduction of the scattering integral with time and to the expected asymptotic behavior of the relaxation dynamics. We do not follow the Boltzmann dynamics here.} 
$$
\mathcal{I}_{k\approx k_F}[\overline{N_k^{\rm NEQ:QP}}] \approx \frac 32 \ U^2 \rho_F^2 \int_{-\infty}^{\infty} d\epsilon_j \epsilon_j 
\left[ \overline{\Delta N^{\rm NEQ}(t, \epsilon_j)}- {\Delta N_k^{\rm EQU}} \right]
= \frac 32 \  U^2 \rho_F^2 \ \overline{\Delta E^{\rm KIN}}
$$
Inserting this into (\ref{QBE}) and using --for not too strong derivations from the adiabatic limit-- the quadratic dependence of the kinetic energy on the ramping time (\ref{EtempDEP}) shows that the corresponding elastic relaxation occurs on a separated time scale
\begin{equation}
\tau_{\rm el} \sim \frac 1{U^2 \rho_F^2 \ \Delta E^{\rm KIN}} \sim \frac {t_R^2}{U^4}
\label{ThermalizationTimeScale}  
\end{equation}
For small values of $U$, this can be much larger than the first relaxation time. Moreover, long ramp-up times delay the onset of elastic relaxation. This is a consequence of the reduced deviations of the transient momentum distribution from its equilibrium counterpart; hence even less phase space is available for scattering processes and thermalization due to Boltzmann dynamics is even more delayed.

In consequence, the temporal separation of the elastic relaxation by scattering events from the earlier energy relaxation spans a transient prethermalization regime. Approaching the adiabatic limit, i.e. for long ramp-up times $t_R$ or weak interactions $U$, this time regime becomes more extended in time. This agrees with the physical intuition on the momentum relaxation by scattering processes which should be enhanced by higher kinetic energies. Such observations parallel those made, for instance, in glasses where relaxation times from transient nonequilibrium states grow with decreasing kinetic energies (low temperatures).

The most important observation is that within the prethermalization regime the kinetic energy does not equal an effective temperature. Instead, the continued presence of an approximate or even sharp \cite{Uhrig2009} discontinuity at the Fermi energy suggests a zero-temperature description for the prethermalized transient state. 
After thermalization has been completed, the kinetic energy determines an effective temperature $\mathcal{T}_{\rm eff}$ defined by $\Delta E^{\rm KIN} = \pi^2 \rho_F \mathcal{T}_{\rm eff}^2/6$. Inserting the temperature dependence of $\Delta E^{\rm KIN}$ (\ref{EtempDEP}) one observes that the effective temperature is proportional to the inverse ramp-up speed. 
\begin{equation}
\mathcal{T}_{\rm eff} \sim \frac{\abs{U}}{t_R} 
\label{TeffUT}
\end{equation}

\section{Consequences for the observability of nonequilibrium BCS dynamics}

As the above results, which have been obtained for the repulsive Hubbard model, depend only on even powers of the interaction they can be carried over to the case of an attractive interaction $U<0$. In both cases the initial state is given by the noninteracting Fermi gas but the final states after the adiabatic or sudden ramping differ. For both attractive and repulsive interactions a correlated many-particle state builds up; however, below the critical temperature $T_c$ a Fermi liquid with pairwise attractive interactions around the Fermi surface exhibits the distinct feature of superconductivity or superfluidity: Then a single particular normal mode of the Fermi liquid becomes unstable with respect to Cooper pairing of low energy fermions. 

The pairing behavior can be discussed both within the attractive Hubbard model  and by referring to the effective BCS Hamiltonian
\begin{equation}
\label{BCSHamiltonian}
H_{\rm int}^{\rm BCS} = \lambda(t) \sum_{p,q} c^{\dagger}_{p \up} c^{\dagger}_{-p \down} c_{-q \down} c_{q \up}
\end{equation}
Although both approaches describe pairing in a similar way, they are not fully equivalent: the BCS interaction can be considered as that subset of the two-particle Hubbard interaction matrix elements which describes the behavior of pairs of fermions with opposite momenta. Other two-particle interaction matrix elements of the Hubbard model are not represented in the BCS Hamiltonian (\ref{BCSHamiltonian}). 

The nonequilibrium behavior of the BCS Hamiltonian (\ref{BCSHamiltonian}) following a nonadiabatic switch of the interaction $\lambda(t)$ has been recently discussed \cite{Barankov2004, Yuzbashyan2005, Yuzbashyan2006b, Yuzbashyan2006, Warner2005}.
There it has been shown that a quantum quench in the effective pairing interaction $\lambda$ leads to the buildup of a paired many-particle state which, furthermore, exhibits characteristic nonequilibrium signatures. For example, temporal oscillations of the order parameter of the superfluid transition, the energy gap parameter $\Delta(t)$, have been observed.

However, experiments in ultracold fermions are well-described by the Hubbard model. While it is experimentally feasible to change the relative strength of the Hubbard interaction in time, it is not possible to do so only for that subset of matrix elements which correspond to the BCS Hamiltonian. Hence the question arises whether the dynamics described by the remaining matrix elements interferes with the predicted nonequilibrium BCS dynamics. 

\subsection{Excitation of nonequilibrium BCS dynamics by finite time ramping of the Hubbard interaction}
In the following we will address one aspect of that question: whether the effects of the additional heat inserted by the feasible quench of the Hubbard interaction as compared to a theoretical quench of the BCS interaction alone already rule out the observation of nonequilibrium BCS dynamics. 

Obviously, the inserted energy depends on the ramping time. In order to observe nonequilibrium BCS dynamics in ultracold Fermi gases, two competing requirements must be met:
\begin{itemize}
\item[(i) ]
Firstly, the ramping must be sufficiently nonadiabatic to excite nonequilibrium behavior. 
\item[(ii) ]
Secondly, the system must not heat up beyond the critical temperature of superconductivity despite the energy intake by the ramping.
\end{itemize}
In a previous work \cite{Barankov2006A} it has been pointed out that a ramp-up which is nonadiabatic on the scale of the BCS instability, i.e. 
\begin{equation} 
t_R \lesssim 1/\Delta
\label{TDelta}
\end{equation}
is sufficient to excite nonequilibrium BCS dynamics. As $\Delta$ is typically a small parameter and as the kinetic energy depends inversely on the ramp-up time, this eases the problem of overheating with respect to a sudden quantum quench. 

\subsection{Overheating of BCS system beyond its critical temperature cannot be excluded}
However, plugging the slowest sufficiently nonadiabatic linear ramp $t_R = t_{\rm BCS} \stackrel{\rm def}= 1/\Delta$ (cf. \ref{TDelta}) into (\ref{TeffUT}) shows that even for that limiting case the energy intake is not negligibly small. Instead, the final temperature is of a similar order of magnitude as the characteristic energies involved in the BCS dynamics, namely $\mathcal{T}_{\rm eff} \sim U \Delta$. Hence, energetic arguments do not imply a clear separation of energy scales related to the nonequilibrium BCS dynamics and temperature, respectively. 
Therefore, we expect that \emph{finally} the temperature will influence and possibly destroy the nonequilibrium BCS behavior.

\subsection{Prethermalization opens window for observation of nonequilibrium BCS behavior}
Fortunately, the thermalization of the momentum distribution is delayed by prethermalization. For the limiting ramping time $t_R=1/\Delta$ thermalization due to elastic scattering processes is deferred to the time scale set by
$$
\tau_{\rm el}  \sim \frac {t_R^2}{U^4} \stackrel{BCS}{\sim} \frac 1{U^4 \Delta^2} \sim \frac 1{U^4} e^{2/\rho_F U} \gg t_{\rm BCS}
$$
For small interaction strength, the elastic relaxation time becomes much larger than the time scale of the BCS dynamics. Hence a long-lasting transient time window is opened by prethermalization; during this time the momentum distribution around the Fermi energy resembles that one of a Fermi liquid at zero temperature and most of the excitation energy is virtually stored in high energy modes sufficiently far apart from it. Thus we expect the buildup of the BCS instability and the persistence of BCS dynamics until quasiparticle scattering redistributes the excitation energies down to low energy modes, implying thermalization.

Based on our observation of prethermalization we are happy to report that  nonequilibrium BCS dynamics is, in principle, accessible by suitable linear ramping protocols performed  with ultracold fermions. Other ramping protocols as discussed by Eckstein and Kollar can further facilitate these experiments by reducing the effective temperature.

\begin{figure}
 \begin{center}
      \includegraphics[width=120mm]{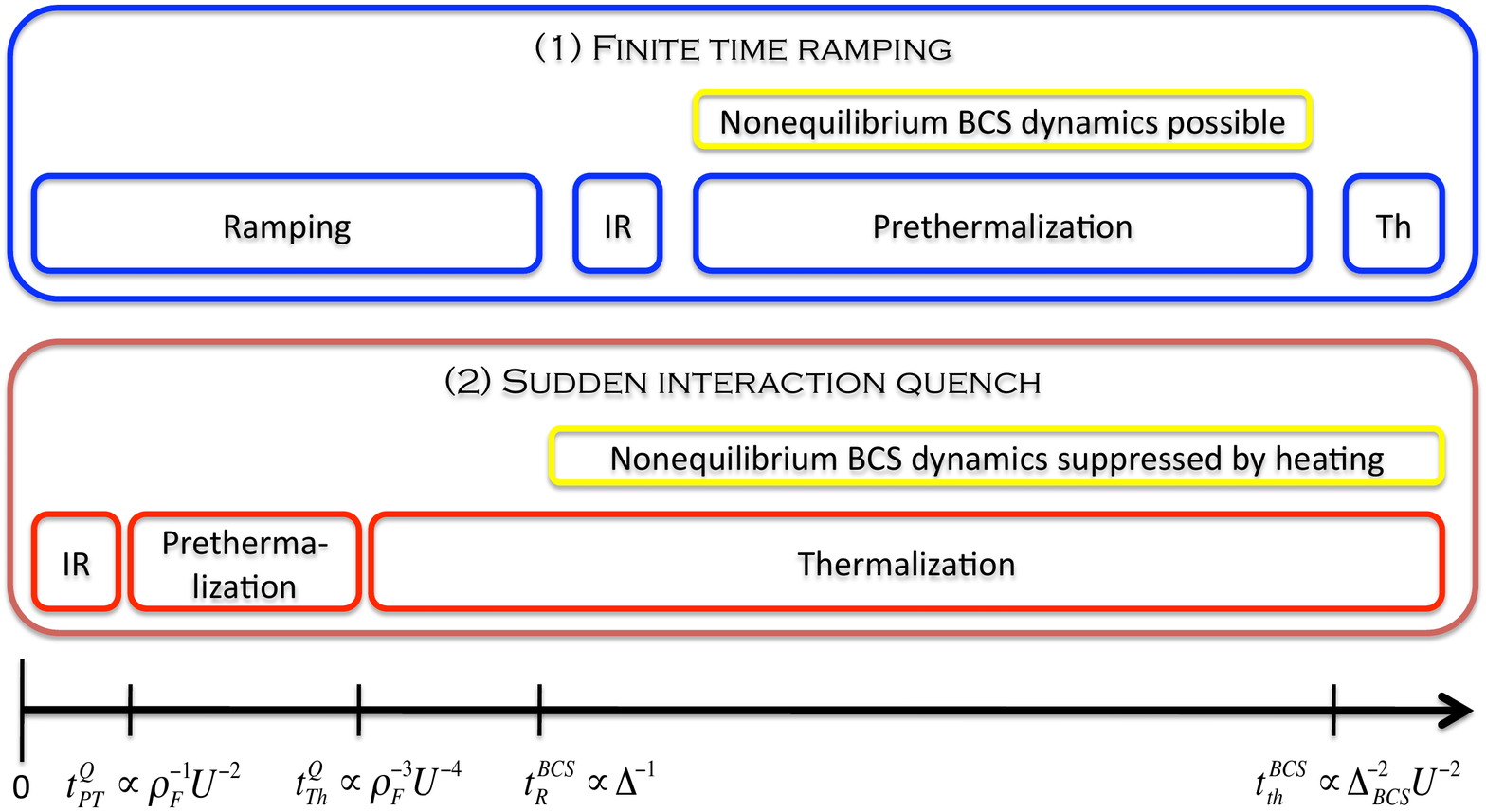}  
      \vspace{-1cm}
    \caption[]{This sketch compares the relevant relaxation time scales in the case of linear finite time ramping on the time scale $t_R \sim 1/\Delta_{\rm BCS}$ (1) with those following a quantum quench (2); the later is assumed sudden w.r.t. the time scale set by the hopping matrix element. In both cases an initial relaxation (IR) is followed by a transient prethermalization regime and final thermalization (Th). The finally achieved effective temperature is of the order or larger than the critical temperature of superconductivity. Note that the prethermalization regime following a sudden quench may not reach into the time regime where signatures of the BCS nonequilibrium dynamics can be expected. Hence the long duration of the prethermalization regime in the case of finite time ramping is crucial for a clear observability of nonequilibrium BCS dynamics.}
      \label{FigLast}
 \end{center}
\end{figure}

\section{Conclusions}
In this paper we have analyzed the implications of linearly ramping up the interaction strength of the Hubbard model in more than one dimension and at zero temperature. We assumed a finite ramping time and discussed corrections to the adiabatic limit as well as the crossover into a quantum quench using second order Keldysh perturbation theory in real time. We observed different relaxation behavior for two observables, namely the momentum distribution and the kinetic energy. On the one hand, the momentum distribution does not relax on a first time scale which is related to second order perturbation theory; it shows a characteristic mismatch of the quasiparticle residue if compared with adiabatic ramping. This mismatch provides a quantitative measure for the nonequilibrium character of the momentum distribution in the first time regime after ramping. 
On the other hand the kinetic energy summed over all momentum modes remains unchanged under the  subsequent Boltzmann dynamics of the momentum distribution; the later becomes effective on a second, much longer time scale and leads to the thermalization of the momentum distribution.  Consequently, a transient time regime characterized by prethermalization appears between these two well-seperated time scales. During this regime the excitation energy of the switching process is present predominantly in high energy momentum modes; around the Fermi energy a zero-temperature description remains applicable. For small deviations from adiabatic switching, i.e. for small energy intakes due to the ramping, this regime is particular extended in time. We compute the dependence of the final temperature on the ramping time and explain how differing expectations have been obtained previously.
Finally we apply our findings to the question whether nonequilibrium BCS dynamics can be observed in ultracold Fermi gases. On the one hand, the final temperature reaches the same order of magnitude as the critical temperature of the BCS problem which makes the observation of such dynamics in a true long-time limit --at least-- very difficult. On the other hand, a suitable choice of the ramping speed and the observed prethermalization of the momentum distribution function open a long transient time window; during that time the excitation energy inserted by the ramping remains located in high energy momentum modes. Therefore, nonequilibrium BCS dynamics, which happens on a small energy scale around the Fermi surface, is not affected by the inserted energy. We expect that prethermalization fosters in a similar way the observability of low-temperature nonequilibrium quantum dynamics of other phenomena which also require the presence of a sharp Fermi surface. 

\ack 
We acknowledge valuable discussions with M. Kollar and B. Altshuler. This work was supported through SFB 631 of the Deutsche Forschungsgemeinschaft, the Center
for NanoScience (CeNS) Munich, and the German Excellence Initiative via
the Nanosystems Initiative Munich (NIM). M.M. acknowledges the support of
the German National Scholarship Foundation.

\appendix
\section{Keldysh perturbation theory for the Hubbard model}
\label{AppendixKeldysh}
In the following appendix we sketch the real-time Keldysh perturbative approach to the Hubbard model. For conventional perturbation theory of the ground state the Gell-Mann and Low theorem links the asymptotically free ingoing ($t\rightarrow -\infty$) and outgoing ($t\rightarrow \infty$) ground state. Interaction effects are switched on adiabatically, treated perturbatively and switched off again on the way of a linear time evolution.  Perturbation theory beyond the ground state, however, cannot rely on the Gell-Mann and Low theorem which only links ground states. Therefore, calculations have to be performed on a closed time contour, mapping the outgoing state back into a representation of the ingoing states. The Keldysh technique \cite{Keldysh1965} has implemented this contour evolution for the calculation of nonequilibrium Greens functions by representing those on a $2\times 2$ Keldysh index space.  Hence, Feynman rules have to be set up which include the matrix structure of Keldysh Greens functions. Details of the following calculation can be found in \cite{MoeckelPhD}. 

\subsection{Nonequilibrium Feynman rules for the Hubbard interaction}
The tensor structure of Feynman rules for the Hubbard interaction on Keldysh space can be constructed in analogy with any exchange interaction: Fundamental vertices describe the emission and absorption of exchange bosons by free fermions; working in the Larkin-Ovchinnikov representation of Keldysh space, their nontrivial tensor structure can be found in \cite{Rammer_NEQQFT} and is quoted in table \ref{Tab1}. A Keldysh representation of the Hubbard interaction can be constructed by linking emission ($\tilde{\gamma}$) and absorption ($\gamma$) vertices by a local, instantaneous boson propagator ($\bf D$) which is diagonal in spin and Keldysh space. The time dependence of the ramping process is included in an explicitly time-dependent interaction strength $U(t)$ and symmetrically attributed to both vertices. In the Larkin-Ovchinnikov representation the Keldysh matrix Greens function can be composed from the retarded ($G^R$), advanced ($G^A$) and Keldysh component ($G^K$) Greens function. 
\begin{equation}
{\bf G}_{\kappa \kappa'}(x, t, \sigma_z, x', t', \sigma_z') = 
\left(\begin{array}{cc}
G^R & G^K \\ 0 & G^A  
\end{array} \right) (x, t, \sigma_z, x', t', \sigma_z')
\label{FFProp}
\end{equation}
For noninteracting fermions, $G^R$, $G^A$ and $G^K$ are given by
\begin{equation}
{\bf G}^{(0)}_{\kappa \kappa'}(x, t, \sigma_z, x', t', \sigma_z') = 
-i  \sum_k e^{ik(x_1-x_2)-i\epsilon_k(t_1-t_2)}
\left(\begin{array}{cc}
\Theta(t_1-t_2) & (1-2n_k) \\ 0 & -\Theta(t_2-t_1)  
\end{array} \right)
\label{FFProp2}
\end{equation} 
This makes the free fermion propagator explicit. 
Diagrams can be evaluated using real-space Feynman rules for Greens functions. Note that whenever propagators are linked to vertices, the related Keldysh indices have to be contracted. 

\begin{table}
\begin{tabular}{lll}
\\
Fermionic propagator:&
\begin{fmffile}{F1}\begin{fmfgraph*}(40,10) \fmfleft{i} \fmfright{o}\fmf{fermion}{i,o} \end{fmfgraph*}\end{fmffile}&
 ${\bf G}_{\kappa \kappa'}(x, t, \sigma_z, x', t', \sigma_z') \qquad \text{(cf. \ref{FFProp})}$ \\ \vspace{2mm}
Boson propagator: &
\begin{fmffile}{F2}\begin{fmfgraph*}(40,10) \fmfleft{i} \fmfright{o}\fmf{boson}{i,o} \end{fmfgraph*}\end{fmffile}&
 ${\bf D}_{\kappa \kappa'}(x, t, \sigma_z, x', t', \sigma_z')=
 \delta(t-t') \delta(x-x') \delta_{\kappa}^{\kappa'}\delta_{\sigma_z}^{\sigma_z'}$ \\
 \vspace{2mm}
Boson absorption vertex: & 
\begin{fmffile}{F3}\begin{fmfgraph*}(40,20) \fmfleft{i} \fmfright{o1,o2}\fmf{fermion}{o1,v,o2} \fmf{boson}{i,v}\end{fmfgraph*}\end{fmffile}&
$\gamma_{\kappa' \kappa''}^{\kappa}(t)=\left( \begin{array}{c} \gamma_{\kappa' \kappa''}^1  \\ \gamma_{\kappa' \kappa''}^2 \end{array} \right)
= \frac{\sqrt {U(t)}}{\sqrt 2}\left( \begin{array}{c} \delta_{\kappa' \kappa''}  \\ \tau_{\kappa' \kappa''}^{(1)} \end{array} \right) $ \\ \vspace{2mm}
Boson emission vertex: &\begin{fmffile}{F4}\begin{fmfgraph*}(40,20) \fmfleft{i1,i2} \fmfright{o}\fmf{fermion}{i1,v,i2} \fmf{boson}{v,o}\end{fmfgraph*}\end{fmffile} &
$\tilde{\gamma}_{\kappa' \kappa''}^{\kappa}(t)=\left( \begin{array}{c} \tilde{\gamma}_{\kappa' \kappa''}^1  \\ \tilde{\gamma}_{\kappa' \kappa''}^2 \end{array} \right)
= \frac{\sqrt {U(t)}}{\sqrt 2}\left( \begin{array}{c} \tau_{\kappa' \kappa''}^{(1)} \\ \delta_{\kappa' \kappa''}  \end{array} \right) $
\end{tabular}
\caption{Feynman rules for the Hubbard interaction in Keldysh space. $\kappa$ represents indices in Keldysh space, $\sigma$ in spin space, $\tau^{(i)}_{\kappa' \kappa''}$ the $2\times2$ Pauli matrices. The boson absorption and emission vertices (which are 3-tensors in Keldysh space) differ in the chosen Larkin-Ovchinnikov representation.}
\label{Tab1}
\end{table}
\vspace{0.5cm}

\subsection{Second order Keldysh perturbation theory for the Hubbard model}
\begin{figure}
\begin{center}
\vspace{0.3cm}
\begin{fmffile}{F7} 
\begin{fmfgraph}(50,40) 
\fmfstraight
\fmfleft{i,ib} 
\fmfright{ob,o} 
\fmftop{ul,ur} 
\fmfbottom{i,ib,x,y,ob,o} 
\fmf{plain}{i,ib} 
\begin{fmfsubgraph}(5,5)(30,30)
\fmf{plain}{ib,x} 
\fmf{fermion}{x,y} 
\fmf{boson}{i,ul}
\fmf{boson}{o,ur}
\fmf{plain,tension=5}{ob,o} 
\fmf{fermion,left=.5}{ul,ur,ul} 
\end{fmfsubgraph}
\fmf{plain}{y,ob} 
\end{fmfgraph} 
\end{fmffile}
\end{center}
\caption{The only diagram contributing to a second order perturbative expansion of the Keldysh Greens function.}  
\label{FigSetSun}
\end{figure}
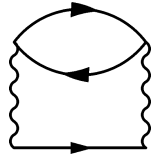

Fortunately, perturbation theory for the Hubbard model is simplified by various symmetries: the bare first-order Hartree diagram is cancelled by a symmetric definition of the Hubbard interaction, renormalized Hartree diagrams vanish to all orders because of particle-hole symmetry and Fock diagrams do not arise because the interaction is effective only between particles of different spin. In consequence,  only a single diagram contributes to a second order perturbative calculation of the Keldysh Greens function (cf. fig \ref{FigSetSun}). 

\subsubsection{Polarization operator.}
In order to make the explicit expressions more handy it is advisable to split off the calculation of the closed fermionic loop, known as the polarization operator~$\bf \Pi$. Moreover, the shorthand notation $i =\lbrace x_i, t_i, \sigma^z_i \rbrace$ where $i$ is an integer number is introduced. To avoid confusion, it does not extend to Keldysh indices. The later are, generically, denoted by Greek letters or, explicitly, by their values (1~or~2). 

\begin{equation}
{\bf \Pi}_{\kappa_3 \kappa_2}(3,2) =
\sum_{\eta_4 \eta_5 \nu_4 \nu_5}
\tilde{\gamma}^{\kappa_3}_{\eta_5 \nu_5}(t_3) 
{\bf G}_{\eta_5 \eta_4}(3,2)  
{\bf G}_{\nu_5 \nu_4}(2,3)  
\gamma^{\kappa_2}_{\eta_4 \nu_4}(t_2)
\end{equation}
Performing the contractions of the Keldysh indices allows to represent the polarization operator as a $2 \times 2$ matrix in Keldysh space.
\begin{equation}
{\bf\Pi}(3,2) =
\frac {\sqrt{U(t_3)U(t_2)}}2 \left(
\begin{array}{c|c}
G^R(3,2)G^K(2,3) & G^A(3,2)G^R(2,3)+G^R(3,2)G^A(2,3)\\
+G^K(3,2)G^A(2,3)&\qquad +G^K(3,2)G^K(2,3) \\
\hline
0 & G^A(3,2)G^K(2,3)+G^K(3,2)G^R(2,3) 
\end{array}
\right)
\label{PolOP}
\end{equation}
This defines naturally the components $\Pi^R=\Pi_{11}$, $\Pi^K=\Pi_{12}$ and $\Pi^A=\Pi_{22}$. 

\subsubsection{The 'setting sun' diagram.}
The second order correction to the Keldysh Greens function follows from the evaluation of the diagram figure \ref{FigSetSun} which can be written as
\begin{multline}
{\bf G}^{(2)}_{\kappa_1 \kappa_{1'}}(1,1') =
\frac 12 \int dx_2 dx_3 dx_4 dx_5 \int dt_2 dt_3 dt_4 dt_5 
\sum_{\kappa_2 \kappa_3 \kappa_4 \kappa_5 \eta_2 \eta_3 \nu_2 \nu_3}
{\bf G}_{\kappa_1 \eta_3}(1,3) 
\gamma^{\kappa_3}_{\eta_3 \nu_3}(t_3)
 \\ \times
{\bf D}_{\kappa_3 \kappa_5}(3,5) 
{\bf \Pi}_{\kappa_5 \kappa_4}(5,4) 
{\bf D}_{\kappa_4 \kappa_2}(4,2)  
{\bf G}_{\nu_3 \nu_2}(3,2)
\tilde{\gamma}^{\kappa_2}_{\nu_2 \eta_2}(t_2) 
{\bf G}_{\eta_2 \kappa_1} (2,1') 
\end{multline}
To achieve a second order correction $\bf G$ denotes on the right hand side the respective noninteracting Greens function (superscripts are suppressed). 
Performing all contractions over Keldysh indices and the integrations over trivial bosonic propagators leads to  
\begin{subequations}
\label{Keldysh2OPTGFallComp}
\begin{align}
\nonumber
{\bf G}^{(2)}_{\kappa_1 \kappa_{1'}}(1,1') =&
\frac 12 \int dx_2 dx_3  \int dt_2 dt_3  \frac {\sqrt{U(t_3)U(t_2)}}2
\\ &\label{Keldysh2OPTGFallCompa}
\hspace{-3mm}\bigg\lbrace 
{G}_{\kappa_11}(1,3) {G}^R(3,2) {\Pi}^R(3,2) {G}_{2 \kappa_1} (2,1') +
\\ & \label{Keldysh2OPTGFallCompb}
{G}_{\kappa_11 }(1,3) {G}^K(3,2) {\Pi}^R(3,2) {G}_{1 \kappa_1} (2,1') + 
\\ & \label{Keldysh2OPTGFallCompc}
{G}_{\kappa_12 }(1,3) {G}^A(3,2) {\Pi}^R(3,2) {G}_{1 \kappa_1} (2,1') +
\\ & \label{Keldysh2OPTGFallCompd}
{G}_{\kappa_11 }(1,3) {G}^A(3,2) {\Pi}^A(3,2) {G}_{2 \kappa_1} (2,1') +
\\ & \label{Keldysh2OPTGFallCompe}
{G}_{\kappa_12 }(1,3) {G}^R(3,2) {\Pi}^A(3,2) {G}_{1 \kappa_1} (2,1') +
\\ & \label{Keldysh2OPTGFallCompf}
{G}_{\kappa_12 }(1,3) {G}^K(3,2) {\Pi}^A(3,2) {G}_{2 \kappa_1} (2,1') +
\\ & \label{Keldysh2OPTGFallCompg}
{G}_{\kappa_11 }(1,3) {G}^R(3,2) {\Pi}^K(3,2) {G}_{1 \kappa_1} (2,1') +
\\ & \label{Keldysh2OPTGFallComph}
{G}_{\kappa_11 }(1,3) {G}^K(3,2) {\Pi}^K(3,2) {G}_{2 \kappa_1} (2,1') +
\\ & \label{Keldysh2OPTGFallCompi}
{G}_{\kappa_12 }(1,3) {G}^A(3,2) {\Pi}^K(3,2) {G}_{2 \kappa_1} (2,1') \bigg\rbrace
\end{align}
\end{subequations}
From this result second order corrections to all components of the Keldysh matrix Greens function can be extracted by specifying the remaining Keldysh indices $\kappa_1$ and $\kappa_1'$. 

\subsection{Second order correction for the Keldysh component Greens function}
Evaluating the Keldysh component Greens function $G^K$ is sufficient to extract the information on the momentum distribution function (cf. \ref{KeldyshMDFwithKGFdef}). This is done by setting $\kappa_1=1$ and $\kappa_{1'}=2$ in (\ref{Keldysh2OPTGFallComp}). 
Inserting (\ref{FFProp2}) together with (\ref{PolOP}) into the Keldysh component of (\ref{Keldysh2OPTGFallComp}) shows that -- because of mismatching time restrictions imposed by the involved $\Theta$-functions -- the terms (\ref{Keldysh2OPTGFallCompc}) and (\ref{Keldysh2OPTGFallCompe}) vanish. The remaining terms can be paired such that all pairs share the same momentum sums and spatial dependences. 
\begin{subequations}
\label{Keldysh2OPTKGF}
\begin{align}
\nonumber
& {\bf G}^{(2)}_{\kappa_1 \kappa_{1'}}(1,1') =
\frac 12 \int dx_2 dx_3  \int dt_2 dt_3  \frac {\sqrt{U(t_3)U(t_2)}}2\bigg\lbrace 
\\ &\label{Keldysh2OPTKGF13}
\tag{\ref{Keldysh2OPTKGF}.1+3}
{G}^R(1,3) {G}^R(3,2) {\Pi}^R(3,2) {G}^A (2,1') + 
{G}^R(1,3) {G}^A(3,2) {\Pi}^A(3,2) {G}^A (2,1') +  
\\ & \label{Keldysh2OPTKGF24}
\tag{\ref{Keldysh2OPTKGF}.2+4}
{G}^R(1,3) {G}^K(3,2) {\Pi}^R(3,2) {G}^K (2,1') + 
{G}^K(1,3) {G}^K(3,2) {\Pi}^A(3,2) {G}^A (2,1') + 
\\ & \label{Keldysh2OPTKGF57}
\tag{\ref{Keldysh2OPTKGF}.5+7}
{G}^R(1,3) {G}^R(3,2) {\Pi}^K(3,2) {G}^K (2,1') + 
{G}^K(1,3) {G}^A(3,2) {\Pi}^K(3,2) {G}^A (2,1') 
\\ & \label{Keldysh2OPTKGF6}
\tag{\ref{Keldysh2OPTKGF}.6}
{G}^{R}(1,3) {G}^K(3,2) {\Pi}^K(3,2) {G}^{A} (2,1') 
\bigg\rbrace
\end{align}
\end{subequations}
Although they also share a common time dependent phase factor $e^{i (t_2-t_3) (\epsilon_p-\epsilon_{p'}+\epsilon_{q_2}-\epsilon_{q_1})}$, their time dependencies $\mathfrak{T}_i$ and phase space factors $\mathfrak{P}_i$ vary.
Integrating out the internal positions makes momentum conservation explicit. 
Then the Keldysh component Greens function can be written as a sum over four pairs
\begin{multline}
\label{KeldyshKGFRSrep}
{\bf G}^{K(2)}(1,1') =
\frac i4 \ U^2  \sum_{p'pq_1q_2}  e^{iq_1(x_1-x_{1'})} e^{-i\epsilon_{q_1} (t_1-t_{1'})} 
\times \\ \times
\sum_{i=1\ldots 4} \mathfrak{T}_i(t_1, t_{1'}, \Delta\epsilon) \mathfrak{P}_i(n_{p'},n_p,n_{q_1},n_{q_2})\delta(p'-p-q_2+q_1)
\end{multline}  
The phase space factors $\mathfrak{P}_i$ can be easily read off
\begin{subequations}
\label{KeldyshPSF}
\begin{align}
\mathfrak{P}_{1+3}(n_{p'},n_p,n_{q_1},n_{q_2}) = & -(n_{p'}-n_p) 
\tag{\ref{KeldyshPSF}.1+3}\\
\mathfrak{P}_{2+4}(n_{p'},n_p,n_{q_1},n_{q_2})= & \
(1-2n_{q_1})(1-2n_{q_2}) (n_{p'}-n_p)
\tag{\ref{KeldyshPSF}.2+4}\\
\mathfrak{P}_{5+7}(n_{p'},n_p,n_{q_1},n_{q_2}) = & \
(1-2n_{q_1}) [n_{p'}+n_p-2n_{p'}n_p]
\tag{\ref{KeldyshPSF}.5+7}\\
\mathfrak{P}_6(n_{p'},n_p,n_{q_1},n_{q_2})  = & 
-(1-2n_{q_2}) [n_{p'}+n_p-2n_{p'}n_p]
\tag{\ref{KeldyshPSF}.6}
\end{align}
\end{subequations}
The time dependent integration kernels $\mathfrak{I}_i$ are evaluated for the linear ramp-up scenario (\ref{DefLinearRamping}), assuming that $t_1, t_{1'}> t_R$. This means, in a strict interpretation, that the obtained result is only applicable to study the behavior of the momentum distribution function as soon as the full interaction has been reached. Transient phenomena during the ramp-up are neglected. Then integrating out internal times leads to
\begin{eqnarray}
\label{KeldyshLinearTDT1324}
\mathfrak{T}_{1+3}   =
\mathfrak{T}_{2+4}   =
\mathfrak{T}_{5+7} &= & \hspace{-5mm}
\frac 1{t_R^2} \frac{2(1-\cos(\Delta\epsilon t_R))}{(\Delta\epsilon)^4} 
\\&\hspace{6mm} +& \nonumber
\frac 1{T} \frac{i}{(\Delta \epsilon)^3} \left[e^{i\Delta\epsilon(t_R-t_1)}-e^{-i\Delta\epsilon(t_R-t_{1'})}+e^{i\Delta\epsilon t_{1'}}-e^{i\Delta\epsilon t_1} \right]  \\&\hspace{6mm} +&
\nonumber
\left[\frac 1{(\Delta\epsilon)^2} + i\frac{t_{1'}-t_1}{\Delta\epsilon} \right]
\\
{\mathfrak{T}_6} &= & \hspace{-5mm}
\label{KeldyshLinearTDT6}
\frac 1{t_R^2} \frac{2(1-\cos(\Delta\epsilon t_R))}{(\Delta\epsilon)^4} 
\\&\hspace{6mm} +& \nonumber
\frac 1{t_R} \frac{i}{(\Delta \epsilon)^3} \left[e^{i\Delta\epsilon(t_R-t_1)}-e^{-i\Delta\epsilon(t_R-t_{1'})}+e^{i\Delta\epsilon t_{1'}}-e^{-i\Delta\epsilon t_1} \right]  \\&\hspace{6mm} +&
\nonumber
\frac 1{(\Delta\epsilon)^2} e^{i\Delta\epsilon(t_{1'}-t_1)}
\end{eqnarray}
Obviously, this calculation has produced a secular term in (\ref{KeldyshLinearTDT1324}) which is proportional to the difference of external times $ t_{1'}-t_1 $. However, this term does not affect the momentum distribution.

\subsection{Second order correction to the momentum distribution}

Since the momentum distribution follows from the Keldysh component Greens function at equal times (\ref{KeldyshMDFwithKGFdef}), the secular terms in its second order correction drop out from the final result. Moreover, this implies that for linear ramping procedures and the assumption $ t_{1'}=t_1 $ all time dependent integration kernels $\mathfrak{T}_{1+3}   =  \mathfrak{T}_{2+4}   = \mathfrak{T}_{5+7} = {\mathfrak{T}_6} =:\mathfrak{T}$ coincide.

Factorizing $\mathfrak{I}$ from the second sum in (\ref{KeldyshKGFRSrep}) allows to add up the phase space factors of the paired terms. In total, this generates 
\begin{equation}
\mathfrak{P}= -4 [n_{q_1}n_{p'}(1-n_p)(1-n_{q_2})-(1-n_{q_1})(1-n_{p'}) n_p n_{q_2}]
\end{equation}
Inserting $\int dE \ \delta( \epsilon_p-\epsilon_{p'}+\epsilon_{q_2}-E)$ and canceling prefactors allows to define the fermionic phase space factor 
\begin{equation}
\label{DefPSFApp}
J_k(E,n) = \sum_{p'pq_2} 
\delta^{p'+q_1}_{p+q_2} \delta(\epsilon_p-\epsilon_{p'}+\epsilon_{q_2}-E)
[n_{q_1}n_{p'}(1-n_p)(1-n_{q_2})-(1-n_{q_1})(1-n_{p'}) n_p n_{q_2}]
\end{equation}
Then reading off the Fourier transform we arrive at the second order correction to the momentum distribution given in (\ref{KeldyshMDFresultKSpace}).

\section{Review of properties of the phase space factor $J_k(E,n)$}
\label{AppendixJ}
\begin{figure}
 \begin{center}
      \includegraphics[width=120mm]{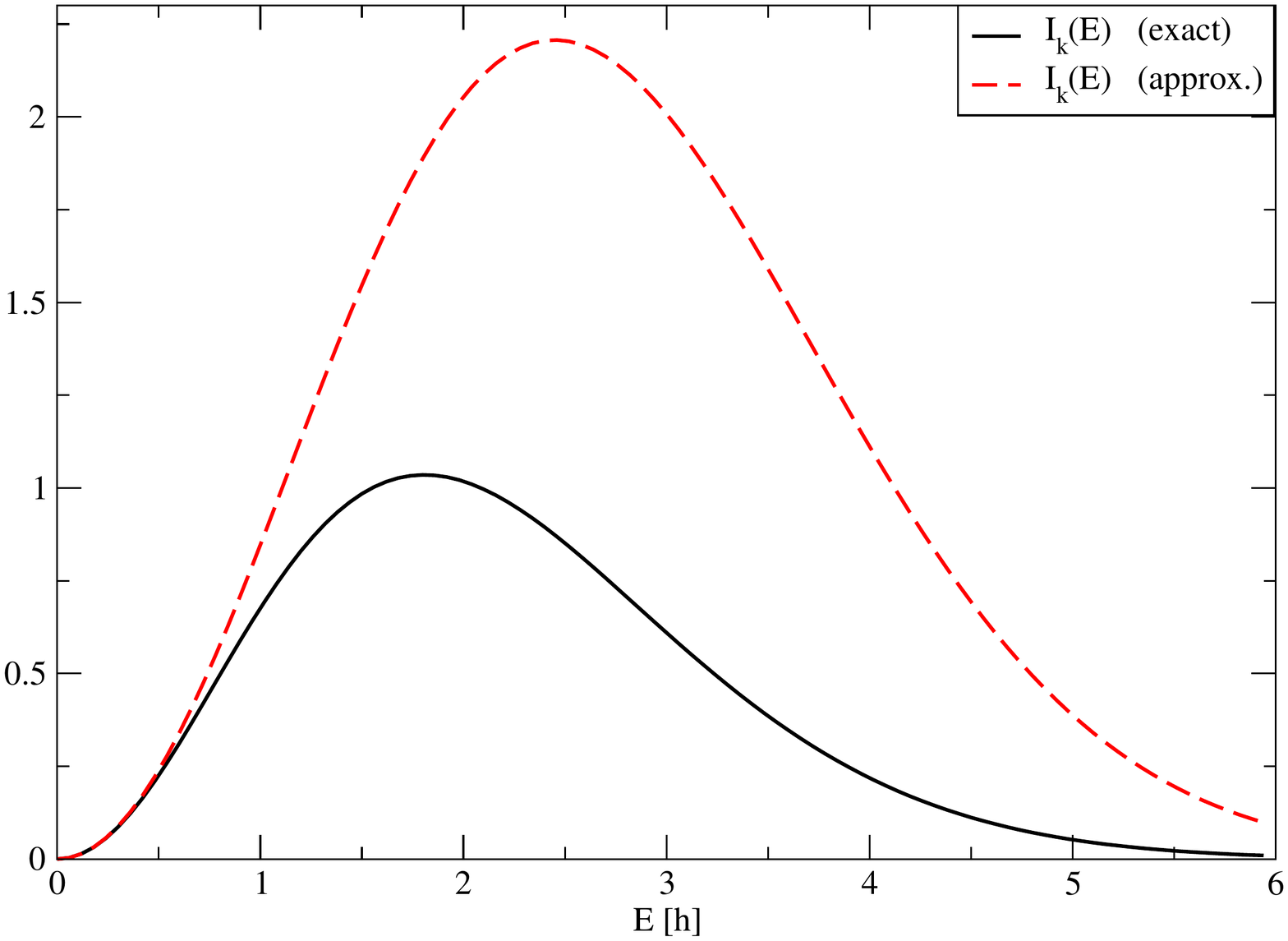}  
    \caption[]{Comparison of the analytical approximate calculation (\ref{PSFApprox}) and a full numerical evaluation of the phase space factor $I_k(E)$. Prefactors are suppressed. One observes that the approximation overestimates the available phase space.}
      \label{Fig7}
 \end{center}
\end{figure}

The fermionic phase space factor $J_k(E,n)$ has been discussed previously \cite{Abrikosov1963,Kehrein_book}. Here we review some of its properties. 
Firstly, it shows quadratic asymptotical behavior for small energies around the Fermi energy $\epsilon_F$.
\begin{equation}
J_k(E,n) \stackrel{E\rightarrow \epsilon_F}{\longrightarrow} \sim (E-\epsilon_F)^2
\end{equation}
Then we denote a second proportionality with $\Theta(x)$ being the Heaviside step function and $I_k(E)$ a suitable reduced phase space factor.
\begin{equation}
J_k(E,n) \sim \left[\Theta(-\epsilon_k) \Theta(E) - \Theta(\epsilon_k) \Theta(-E) \right] I_k(\abs{E})
\quad \Leftrightarrow \quad \left\lbrace 
\begin{array}{cc}
J_k(E<0)=0 & \forall \epsilon_k < 0 \\
J_k(E>0)=0 & \forall \epsilon_k > 0 \end{array} \right.
\end{equation}
This implies that the phase space factor cuts off \emph{all} formal energy divergences in (\ref{KeldyshMDFresultKSpace}). 
Thirdly, we evaluate the reduced phase space factor in the limit of infinite dimensions.
In this limit all three momentum sums in (\ref{DefPSFApp}) are promoted to energy integrations with respect to a Gaussian density of states (\ref{GaussianDoS}). This already includes momentum conservation \cite{Vollhardt1992}. While the first integration is trivial (energy conservation), the second one can be performed analytically. This gives
\begin{equation}
\label{NumStartPSF}
I_k(E) = \left( \frac 1{\sqrt{2\pi h^*}} \right)^3 h^* \ \frac{\pi}2 \ e^{-\frac{E^2}{4h^{*^2}}}
\int_0^E d\epsilon \  e^{-\frac{3\epsilon^2-2E\epsilon}{4h^{*^2}}} 
\left[ \text{erf}\left(\frac{\epsilon}{h^*}\right) - \text{erf}\left(\frac{\epsilon-E}{h^*}\right) \right]
\end{equation}
A first order expansion of the error function effectively decouples two internal energy integrations such that an analytic approximation becomes possible
\begin{equation}
\label{PSFApprox}
I_k(E) \approx \left( \frac 1{\sqrt{2\pi h^*}} \right)^3 E^2 \ e^{-\frac{E^2}{6h^{*^2}}}
\end{equation}
Fig \ref{Fig7} compares this approximation with a full numerical evaluation of (\ref{NumStartPSF}). As for all numerical computations, the bandwidth has been set to six times the hopping rate and an energy resolution of $N=6000$ sites has been used. 

\bibliographystyle{unsrt}

\end{document}